\documentclass[aps,prl,showpacs,twocolumn,superscriptaddress,nofootinbib,floatfix,mamsmath,10pt]{revtex4}
\usepackage{epsfig,epsf,graphics,psfrag}
\usepackage{times}
\usepackage{float}
\usepackage{color}
\usepackage{amsfonts,amssymb,stmaryrd,latexsym,amsmath}

\newcommand{\be}{\begin{equation}}
\newcommand{\ee}{\end{equation}}
\newcommand{\ba}{\begin{eqnarray}}
\newcommand{\ea}{\end{eqnarray}}
\newcommand{\non}{\nonumber}

\newcommand{\al}{&}
\newcommand{\Lag}{\mathcal{L}}
\newcommand{\Amp}{\mathcal{A}}
\newcommand{\Tr}{\textrm{Tr}}
\newcommand{\order}[2][p]{\mathcal{O}(#1^{#2})}

\begin{document}
\title{Examining coupled-channel effects in radiative charmonium transitions}

\author{Feng-Kun Guo}
\email{fkguo@hiskp.uni-bonn.de}
\affiliation{Helmholtz-Institut f\"ur Strahlen- und
             Kernphysik and Bethe Center for Theoretical Physics, \\
             Universit\"at Bonn,  D--53115 Bonn, Germany}

\author{Ulf-G. Mei{\ss}ner}
\email{meissner@hiskp.uni-bonn.de}
\affiliation{Helmholtz-Institut f\"ur Strahlen- und
             Kernphysik and Bethe Center for Theoretical Physics, \\
             Universit\"at Bonn,  D--53115 Bonn, Germany}
\affiliation{Institute for Advanced Simulation, Institut f\"{u}r Kernphysik
             and J\"ulich Center for Hadron Physics, \\
             Forschungszentrum J\"{u}lich, D--52425 J\"{u}lich, Germany}

\begin{abstract}
\noindent Coupled-channel effects due to coupling of charmonia to the charmed and
anticharmed mesons are of current interest in heavy quarkonium physics. However,
the effects have not been unambiguously established. In this Letter, a clean method
is proposed in order to examine the coupled-channel effects in charmonium
transitions. We show that the hindered M1 radiative transitions from the $2P$ to
$1P$ charmonia are suitable for this purpose. We suggest to measure one or more of
the ratios $\Gamma(h_c'\to \chi_{cJ}\gamma)/\Gamma(\chi_{cJ}'\to\chi_{cJ}\pi^0)$
and $\Gamma(\chi_{cJ}'\to h_c\gamma)/\Gamma(\chi_{cJ}'\to\chi_{cJ}\pi^0)$, for
which highly nontrivial and parameter-free predictions are given. The picture can
also be tested using both unquenched and quenched lattice calculations.
\end{abstract}


\maketitle

Thanks to various experiments world-wide, our knowledge of the physics of heavy quarkonium
has been greatly enriched in the last decade. New charmonium(-like) states, including the
$h_c$ and $\chi_{c2}'$ as well as the so-called $XYZ$ states, were observed. Most
of the $XYZ$ states are above the open-charm thresholds, and do not fit the
expectations from the quark model. Hence, it is of current interest and high
importance to investigate the coupled-channel effects, originating from the
coupling of $c\bar c$ to charmed-meson--anticharmed-meson channels, in charmonium
physics. So far, these effects have not been established unambiguously, though
evidences exist, see, e.g., Refs.~\cite{Guo:2010ak,Li:2007xr,Eichten:QWG2011} for
transitions between charmonia
and Refs.~\cite{Eichten:2005ga,Pennington:2007xr}
for spectroscopy. This Letter is devoted to a clean way of examining the
coupled-channel effects. For this purpose, we propose to measure
one or more of the ratios $\Gamma(h_c'\to
\chi_{cJ}\gamma)/\Gamma(\chi_{cJ}'\to\chi_{cJ}\pi^0)$ and $\Gamma(\chi_{cJ}'\to
h_c\gamma)/\Gamma(\chi_{cJ}'\to\chi_{cJ}\pi^0)$, for which highly nontrivial
predictions will be made.

At the hadronic level, one may consider the coupling of heavy quarkonium states to
open-flavor meson and anti\-mesons using effective Lagrangians, and take into account
the coupled-channel effects by calculating intermediate heavy meson loops. Because
the difference between the mass of a heavy quarkonium and heavy meson--antimeson
thresholds is small, the intermediate heavy mesons are nonrelativistic with
velocity $v\ll1$. Based on this observation, a nonrelativistic effective field
theory (NREFT) was proposed~\cite{Guo:2009wr,Guo:2010zk,Guo:2010ak}. It was found
that the transitions between two $P$-wave charmonia with the emission of one pion
are completely dominated by the coupled-channel effects with an enhancement of
$\sim 1/v^3$ in the decay amplitude in contrast to the multipole
contribution~\cite{Guo:2010ak}. Here we find that the hindered M1 transitions from
a $2P$ to a $1P$ charmonium is also dominated by the coupled-channel effects.
Because these two types of transitions cannot be directly connected to each other,
only when both of them are dominated by the coupled-channel effects, nontrivial
predictions can be made in the framework of NREFT. Consequently, the
measurements of hindered M1 transitions from $P$-wave charmonia provides a good
opportunity for examining the coupled-channel effects in charmonium transitions.

\begin{figure}[t]
    \centering
    \includegraphics[width=0.5\textwidth]{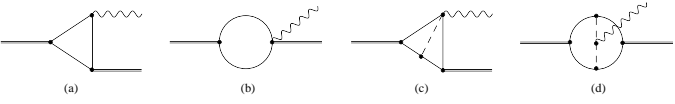}
    \vglue-1mm
    \caption{Possible triangle (a) and two-point (b) loops for the radiative transitions.
    (c) and (d) are two typical two-loop diagrams.
    The double, solid, wavy and dashed lines
    represent charmonia, charmed mesons, photons, and pion, respectively.\label{fig:loop}}
    \end{figure}

There are several nice features of the hindered M1 transitions of $P$-wave
charmonia for investigating the coupled-channel effects of open-charm mesons:\\
1) First of all, these transitions are expected to be dominated by the
    coupled-channel
    effects. On one hand, in quark models, the decay amplitude for an M1
    transition between two heavy quarkonia is proportional to the overlap of
    the wave functions of the initial and final quarkonia (see,
    e.g.~\cite{Barnes:2005pb}),
    \be%
    \Gamma_{\rm M1} \propto |\langle\psi_f|\psi_i\rangle|^2 E_\gamma^3,
    \label{eq:AM1qm}
    \ee%
    with $\psi_{i(f)}$ being the wave function of the initial (final) heavy
    quarkonium, and $E_\gamma$ the photon energy in the rest frame of the
    initial particle. For transitions between a $2P$ and a $1P$ state, if the
    charmonia are purely $c\bar c$ states, the overlap is nonzero only because
    of small relativistic corrections. The statement can be made
    model-independently using the potential nonrelativistic
    QCD~\cite{Brambilla:2004jw} --- the leading contribution
    vanishes~\cite{Brambilla:2005zw}. Hence, the transition amplitude would
    start from $E_\gamma v_c/m_c$, where $v_c$ and $m_c$ are the charm quark
    velocity and mass, respectively, and the factor $1/m_c$ accounts for the
    spin-flip. Indeed, the transition rates are very small in quark model
    calculations
    --- at the largest of the order 1~keV~\cite{Barnes:2005pb}.
    On the other hand,
    because the leading coupling of a $P$-wave charmonium to a charmed-meson
    and anticharmed-meson pair is in an $S$-wave, the decay amplitude through
    intermediate charmed-meson loops as shown in Fig.~\ref{fig:loop}~(a) scales
    as
    \be%
    \label{eq:Aa} \Amp_{\rm (a)} \sim \frac{v^5}{(v^2)^3} \frac{E_\gamma}{m_c}
    = \frac{E_\gamma}{m_c v},
    \ee%
    where $v^5$ and $(v^2)^3$ account for the nonrelativistic loop measure and
    three nonrelativistic propagators, respectively, and $E_\gamma$ comes from
    the $P$-wave coupling of the photon to charmed mesons. The factor of
    $1/m_{c}$ is again due to the spin-flip (more details will be given
    below Eq.~(\ref{eq:Amp})).
    In addition, the amplitude is proportional to the electric charge $e$ and
    the product of coupling constants of the $1P$ and $2P$ states to the
    charmed mesons. Yet, there is no suppression analogous to
    $|\langle\psi_f|\psi_i\rangle|^2$ because the initial and final charmonia,
    though having different principal quantum numbers, do not couple to each
    other directly. Instead, they couple through intermediate charmed mesons,
    and there is no similar suppression for such couplings.
    Note that $v$ in Eq.~(\ref{eq:Aa}) will not approach 0 even when the charmonium mass
    overlaps with the charmed-meson threshold, since it should be
    understood as the average of two velocities corresponding to the two cuts
    in the three-point loop.\\
2) The triangle hadronic loops involved in the transitions are
    convergent in the nonrelativistic framework. Therefore, we do not need to
    introduce a counterterm. On the contrary, similar loops for the M1
    transitions of $S$-wave charmonia are divergent so that similar statements
    cannot be made there. This is another nice example of the important role of
    such diagrams in hadron physics, see e.g. the classical work on neutral
    pion photoproduction off nucleons \cite{Bernard:1991rt} or the more recent
    investigation of the large isospin violation in the decay
    $\eta(1405/1475)\to 3\pi$~\cite{Wu:2011yx}.\\
3) Because the leading coupling of charmed and anticharmed mesons to the
    $P$-wave charmonium is in an $S$-wave, there is no derivative in such
    vertices. Hence the two-point loop with the four-particle contact term
    $\chi_{cJ}D^{(*)}\bar D^{(*)}\gamma$, as shown in Fig.~\ref{fig:loop}~(b),
    is not related to that in Fig.~\ref{fig:loop}~(a) by gauge symmetry. It can
    be treated separately. Being gauge-invariant by itself, the four-particle
    contact term should contain the electromagnetic field strength $F^{\mu\nu}$
    for the photon. Hence, the corresponding vertex is proportional to the
    external momentum of the photon. The amplitude for the diagram scales as
    \be%
    \label{eq:Ab} \Amp_{\rm (b)} \sim \frac{v^5}{(v^2)^2}  \frac{E_\gamma}{m_c}
    = v  \frac{E_\gamma}{m_c}.
    \ee%
    One sees that it is two orders higher in the meson velocity counting than
    the diagram Fig.~\ref{fig:loop}~(a), and hence can be neglected at leading
    one-loop order.\\
4) As will be shown later, the two-loop diagrams (c) and (d) are also
    suppressed compared with (a).

The coupling of the $P$-wave charmonia to the charmed and anticharmed mesons is
described by the Lagrangian~\cite{Colangelo:2003sa}
\be%
{\cal L}_\chi = i \frac{g_1}{2} \Tr\left[ \chi^{\dag i} H_a \sigma^i {\bar
H}_a\right] + {\rm h.c.}, \label{eq:Lchi0}
\ee%
where $g_1$ is the coupling constant of the $1P$ charmonium states ($g_1'$ will be
used for the $2P$ states), $H_a=\vec{V}_a\cdot\vec{\sigma}+P_a$ and ${\bar
H}_a=-\vec{{\bar V}}_a\cdot\vec{\sigma}+{\bar P}_a$ are fields annihilating charmed
and anticharmed mesons, respectively, with $\vec{\sigma}$ the Pauli matrices and
$a$ the light flavor index. The two-component notation
introduced in Ref.~\cite{Hu:2005gf} is used here, which is convenient for processes
with negligible recoil effect (less than 1\% for the processes considered in this
Letter). The $P$-wave charmonia are collected in the spin-multiplet
\be%
\label{eq:chicJ} \chi^i = \sigma^j
\left(-\chi_{c2}^{ij}-\frac{1}{\sqrt{2}}\epsilon^{ijk}\chi_{c1}^k +
\frac{1}{\sqrt{3}}\delta^{ij}\chi_{c0} \right) + h_c^i.
\ee%
The magnetic coupling of the photon to heavy mesons is described by the
Lagrangian~\cite{Amundson:1992yp,Hu:2005gf}
\be%
\label{eq:Lem} \Lag_{\gamma} = \frac{e\beta}{2} \Tr\left[ H_a^\dag H_b^{}
\vec{\sigma}\cdot \vec{B} Q_{ab} \right] + \frac{e Q'}{2m_Q} \Tr \left[ H_a^\dag
\vec{\sigma}\cdot \vec{B} H_a^{} \right],
\ee%
where $B^k=\epsilon^{ijk}\partial^iA^j$ is the magnetic field, $Q={\rm
diag}\{2/3,-1/3,-1/3\}$ is the light quark charge matrix, and $Q'$ is the
heavy quark electric charge (in units of $e$). The first
term describes the nonperturbative physics of the light quarks, while the second
term is for the magnetic coupling of the heavy mesons and hence is proportional to
$1/m_Q$. Although the photon can also couple to the heavy mesons through gauging
the kinetic energy term, this vertex does not contribute to the magnetic transitions.

\begin{table}[h!]
\begin{center}
\renewcommand{\arraystretch}{1.2}
\begin{ruledtabular}
\begin{tabular}{l c }
$h_c'\to\gamma\chi_{c0}$ & $[D,{\bar D}^*,D^*]$, $[D^*,{\bar D},D]$, $[D^*,{\bar D}^*,D^*]$\\
$h_c'\to\gamma\chi_{c1}$ & $[D,{\bar D}^*,D]$, $[D^*,{\bar D},D^*]$, $[D^*,{\bar D}^*,D]$\\
$h_c'\to\gamma\chi_{c2}$ & $[D,{\bar D}^*,D^*]$, $[D^*,{\bar D}^*,D^*]$\\
$\chi_{c2}'\to\gamma h_c$ & $[D^*,{\bar D}^*,D]$, $[D^*,{\bar D}^*,D^*]$\\
\end{tabular}
\end{ruledtabular}
\vglue-1mm \caption{\label{tab:loops}Possible loops contributing to each
transition. The charge-conjugated ones and the flavor labels are not shown for
simplicity.}
\end{center}
\end{table}

Because the $\chi_{cJ}$ states are easier to be detected than the $h_c$, and
$\chi_{c2}'$ has been observed, we will calculate the decay widths of the hindered
M1 transitions $h_c'\to\gamma\chi_{cJ}$ and $\chi_{c2}'\to\gamma h_c$. Results for
the other hindered M1 transitions of the $P$-wave charmonia can be easily obtained
using the same method. Denoting the charmed meson connecting the initial charmonium
and the photon as $M1$, the one connecting two charmonia as $M2$, and the other as
$M3$, we specify the triangle loops by $[M1,M2,M3]$. Considering both the
pseudoscalar and vector charmed mesons, possible loops for these transitions are
listed in Table~\ref{tab:loops} (see also Ref.~\cite{Guo:2010ak} for details).

The decay amplitude for each transition can be expressed in terms of the scalar
three-point loop function \vspace{-5mm}
\begin{widetext}
\ba%
I(q) \equiv i\int\!\frac{d^4l}{(2\pi)^4} 
\frac{1}{\left(l^2-m_1^2+i\epsilon\right)
\left[(P-l)^2-m_2^2+i\epsilon\right] \left[(l-q)^2-m_3^2+i\epsilon\right] },
\ea%
where $P$ and $q$ are the momenta of the initial particle and the photon,
respectively, $m_i(i=1,2,3)$ are the masses of the particles $Mi$ in the loop. The
analytic expression can be found in Refs.~\cite{Guo:2010ak,Cleven:2011gp}. The
amplitude for the transition $\chi_{c2}'\to\gamma h_c$ reads
\ba%
\label{eq:Amp} \Amp(\chi_{c2}'\to\gamma h_c) \al=\al \frac{4ie g_1 g_1'}{3}
   \epsilon^{ijk} \varepsilon^{kl}(\chi_{c2}') \bigg\{
    q^i\varepsilon^j(\gamma)\varepsilon^l(h_c) \left[
    -\left(\beta+\frac{4}{m_c}\right) I(q,D^*,D^*,D)  +
    \left(\beta-\frac{2}{m_c}\right) I(q,D_s^*,D_s^*,D_s)\right] \non\\
    &+& \left[ q^i\varepsilon^j(h_c)\varepsilon^l(\gamma) +
    q^l\varepsilon^i(h_c)\varepsilon^j(\gamma) \right] \left[
    \left(\beta-\frac{4}{m_c}\right) I(q,D^*,D^*,D^*)   -
    \left(\beta+\frac{2}{m_c}\right) I(q,D_s^*,D_s^*,D_s^*) \right]  \bigg\},
\ea%
\end{widetext}
where the loop function has been written as $I(q,M1,M2,M3)$. The charge-conjugated
channels are taken into account.\footnote{\label{foot1}They were not considered in
Ref.~\cite{Guo:2010ak,Guo:2009wr,Guo:2010zk,Guo:2010ca}. Hence all the loop
amplitudes therein should be doubled, and the decay widths from the loops should be
multiplied by 4. All ratios remain innocent. We thank T.~Mehen and D.-L.~Yang for
pointing out this.} The amplitudes for the other transitions can be obtained
similarly.

Since the spin direction of the $c$ or $\bar c$ quark should be flipped in the M1
transitions, the decay amplitude should vanish in the heavy quark limit. It is
nonzero only because of the $\order[m_c]{-1}$ spin symmetry breaking effect. One
easily sees there must be nonvanishing contributions from the second term in the
Lagrangian Eq.~(\ref{eq:Lem}). In fact, the first term, to be called $\beta$-term
in the following, also contributes at the same order though $m_c^{-1}$ is not
explicit in the amplitude. Let us look at the decay amplitude given in
Eq.~(\ref{eq:Amp}). The $\beta$-term contribution would vanish were spin symmetry a
good symmetry, i.e., different loops proportional to $\beta$ cancel each other
exactly if the hyperfine splitting between vector and pseudoscalar charmed mesons
$M_{D_{(s)}}-M_{D_{(s)}^*}$ is tuned to zero. The surviving part is due to the
nonvanishing hyperfine splitting which is of order $m_c^{-1}$.

Because the expansion parameter in the NREFT $v\simeq0.4$ is not small, the results
should have sizeable uncertainties. This can be seen by analyzing the power
counting of the decay amplitudes for certain two-loop diagrams. In
Ref.~\cite{Cleven:2011gp}, it is argued that vertex corrections due to
pion-exchange is suppressed, so that the largest two-loop contribution comes from
diagrams shown in Fig.~\ref{fig:loop}~(c) and (d). The diagram (c) contains four
nonrelativistic charmed meson propagators and one relativistic pion propagator.
Each momentum is of order $M_D v$, so that each propagator scales as $1/v^2$ in the
velocity counting. The photon vertex comes from gauging the charmed-meson--pion
axial coupling, hence it contributes a factor of $g/F_\pi$ with $g$ and $F_\pi$
being the axial coupling constant and pion decay constant, respectively. The
charmed-meson--pion axial coupling is in a $P$ wave. Because this is the only
$P$-wave vertex in the diagram, it should scales as the photon momentum, and the
vertex is proportional to $E_\gamma g/F_\pi$. Therefore, the decay amplitude for
the diagram shown in Fig.~\ref{fig:loop}~(c) scales as
\be%
\label{eq:Ac} \Amp_{\rm (c)} \sim \frac{(v^5)^2}{(v^2)^5}
\frac{g^2}{(4\pi)^2F_\pi^2}  \frac{E_\gamma}{m_c} M_D^2 = \frac{E_\gamma}{m_c}
\left(\frac{g M_D}{\Lambda_\chi}\right)^2,
\ee%
where the factor $1/(4\pi)^2$ appears because there is one more loop than in the
one-loop case, and the chiral symmetry breaking scale is $\Lambda_\chi=4\pi F_\pi$.
In order to compare with Eq.~(\ref{eq:Aa}), a factor of $M_D^2$ is introduced, with
$M_D$ being the charmed-meson mass, to make the whole scaling have the same
dimension as that in Eq.~(\ref{eq:Aa}). The value of the axial coupling constant
$g=0.6$ can be determined from $\Gamma(D^*\to D\pi)$~\cite{Burdman:1992gh,PDG2010}.
Numerically, one has $g M_D/\Lambda_\chi\simeq1$. The diagram (d) has the same
scaling as (c). This can be seen easily because the one more propagator in (d) is
balanced by two more $P$-wave vertices. Therefore, the two-loop diagrams are
effectively suppressed compared with Fig.~\ref{fig:loop}~(a) by a factor of
$v\simeq0.4$.

\begin{figure}[t]
\centering
\includegraphics[width=0.24\textwidth]{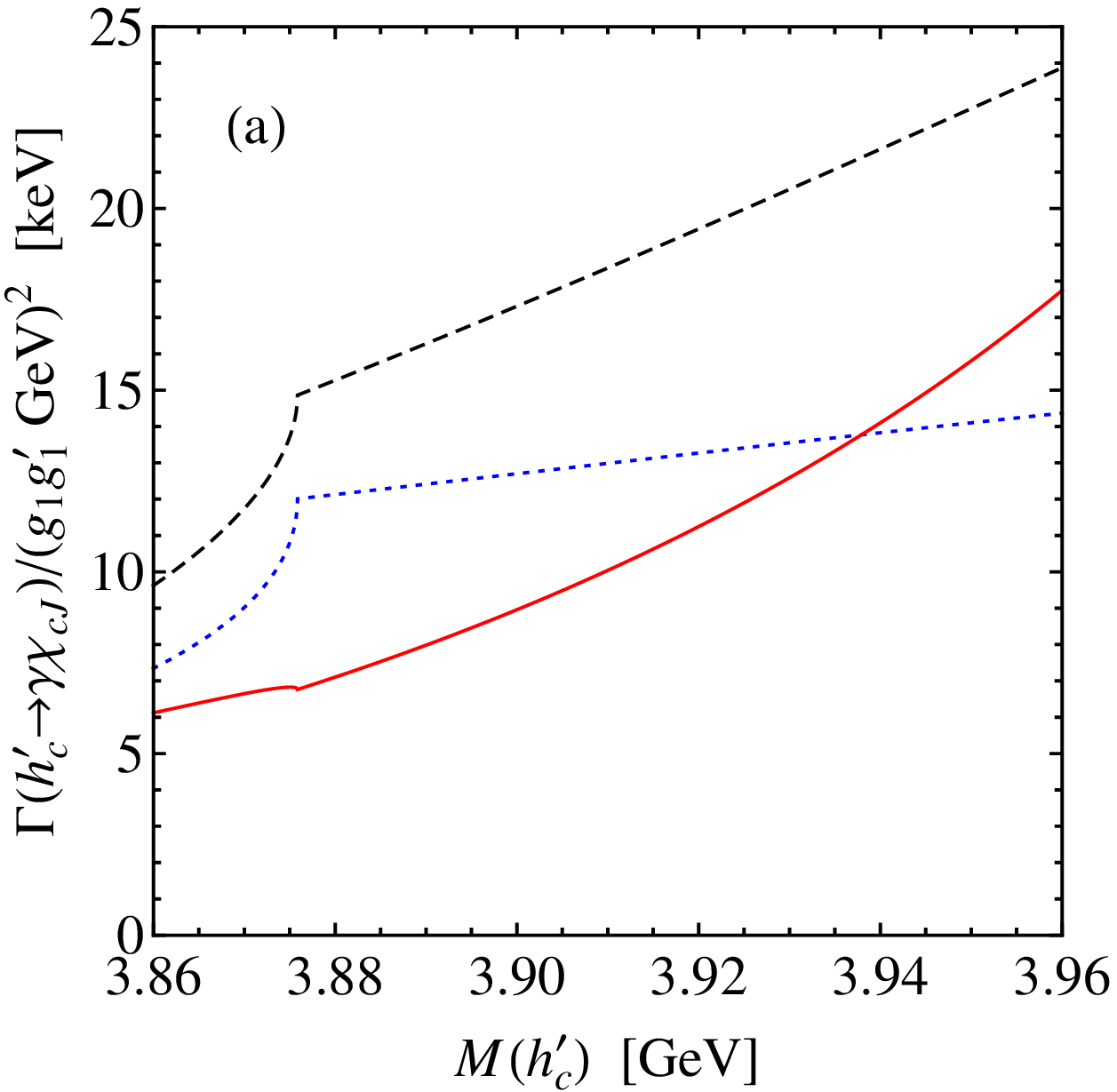} \hglue-1mm
\includegraphics[width=0.24\textwidth]{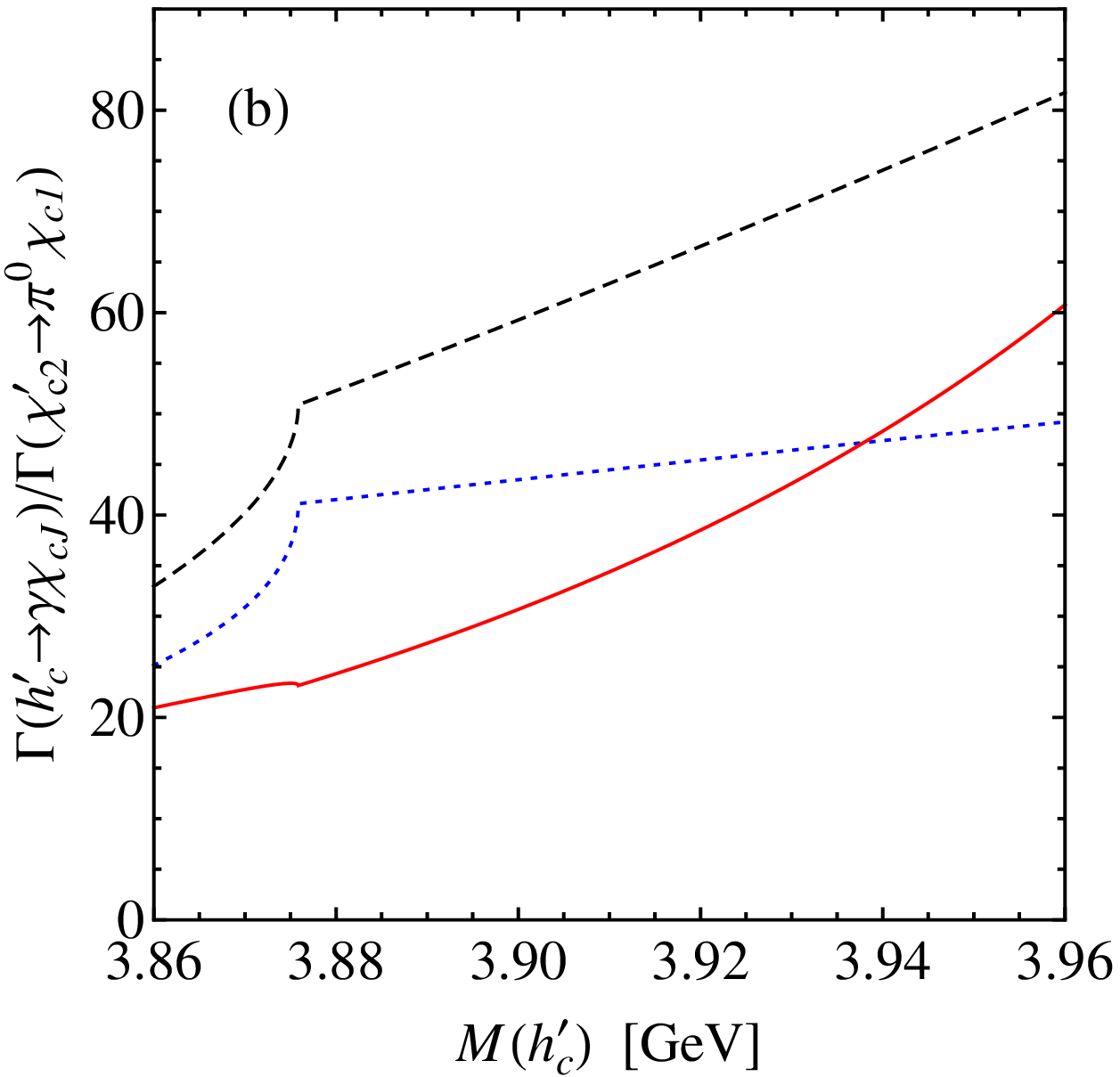}
\caption{(a) Decay widths of the M1 transitions $h_c'\to\gamma\chi_{cJ}$.
(b) Parameter-free ratios $\Gamma(h_c'\to\gamma\chi_{cJ})/\Gamma(\chi_{c2}'\to\pi^0\chi_{c1})$.
The dot-dashed, solid and dashed lines are for $h_c'$ decays with $\chi_{c0}$,
$\chi_{c1}$ and $\chi_{c2}$ in the final state, respectively.
\label{fig:hcM1}}
\end{figure}

In numerical calculations, we use the central values of all measured
masses~\cite{PDG2010}. The value of $\beta$ is not precisely known. Here, we take
the value $\beta^{-1}=276$~MeV determined with $m_c=1.5$~GeV in
Ref.~\cite{Hu:2005gf}. In fact, the precise value of $\beta$ is not important. A
change of $\beta^{-1}$ from 276 to 376~MeV only causes a change in the decay width
of less than 10\%. Hence, the decay width for the $\chi_{c2}'\to\gamma h_c$ is
\be%
\label{eq:width} \Gamma(\chi_{c2}'\to\gamma h_c) = (10.7\pm4.3) \frac{(g_1
g_1')^2}{{\rm GeV}^{-2}}~{\rm keV},
\ee%
where a 40\% uncertainty has been assigned to account for higher order effects.
Because the $\chi_{cJ}$ and $h_c$ are below the open charm thresholds, the coupling
constant $g_1$ cannot be measured directly through the decays of the $P$-wave
charmonia. Similarly, since the only established $2P$ charmonium $\chi_{c2}'$ is
below the $D^*\bar D^*$ threshold, $g_1'$ is also not known yet. In order to obtain
an order-of-magnitude estimate of the branching fraction for the decay
$\chi_{c2}'\to\gamma h_c$, we take a model value
$g_1\simeq-4$~GeV$^{-1/2}$~\cite{Colangelo:2003sa}. \footnote{The value of $g_1$ as
defined in Eq.~(\ref{eq:Lchi0}) is twice of that in~\cite{Colangelo:2003sa}.} For
$g_1'$, we resort to quark model calculations of the decay widths of $\chi_{cJ}'$
and $h_c'$. From the results in the nonrelativistic potential model in
Ref.~\cite{Barnes:2005pb}, we get $g_1'=0.5...1.3$~GeV$^{-1/2}$. From the results
in the Cornell coupled-channel model~\cite{Eichten:2005ga}, we get
$g_1'=0.7...1.2$~GeV$^{-1/2}$. Hence, one may take $g_1'\simeq1$~GeV$^{-1/2}$ as an
estimate. Thus, we get the estimate $\Gamma(\chi_{c2}'\to\gamma h_c)=
\mathcal{O}(170~{\rm keV})$, which is much larger than the quark model prediction
1.3~keV~\cite{Barnes:2005pb}. This is consistent with the analysis made  above that
the transitions are dominated by the coupled-channel effects. The width of the
$\chi_{c2}'$ was measured to be $(24 \pm 6)$ MeV~\cite{PDG2010}. Hence, the
branching fraction for the M1 transition $\chi_{c2}'\to\gamma h_c$ is
\be%
{\cal B} (\chi_{c2}'\to\gamma h_c) =\mathcal{O}(1\times10^{-2}).
\ee%
A future observation with comparable branching fraction would strongly indicate the
dominance of coupled-channel effects in the transition. Similar predictions can be
made for the other M1 transitions of $P$-wave charmonia. The results for the widths
of the transitions $h_c'\to\gamma\chi_{cJ}$ are shown in Fig.~\ref{fig:hcM1}~(a) as
a function of the unknown mass of the $h_c'$. They should be understood to have an
uncertainty of about 40\%. One sees that the partial widths of these transitions
are of the same order as $\Gamma(\chi_{c2}'\to\gamma h_c)$, and are typically of
$\mathcal{O}(100)$~keV if using the same model estimates of $g_1$ and $g_1'$ as
above, much larger than the results in the potential model~\cite{Barnes:2005pb}.

More interestingly, nontrivial parameter-free predictions can be made for the
ratios of the partial widths of these  hindered M1 transitions to those of the
transitions between two $P$-wave charmonia with emission of one pion. In
Ref.~\cite{Guo:2010ak}, it is shown that the latter transitions are also dominated
by the charmed meson loops. The one-loop and largest two-loop diagrams scale as
$q_\pi\Delta/v^3$~\cite{Guo:2010ak} and $(q_\pi\Delta/v^2) (M_D
E_\pi/\Lambda_\chi^2)$~\cite{Cleven:2011gp}, respectively, where $q_\pi (E_\pi)$ is
the pion momentum (energy), and the charged and neutral charmed meson difference
$\Delta$ describes the isospin breaking. One sees that the one-loop diagrams
dominate over the two-loop ones. The decay amplitudes and decay widths for the
single-pion transitions have been calculated in Ref.~\cite{Guo:2010ak}. Here we
only compare the M1 transitions with the decay $\chi_{c2}'\to\chi_{c1}\pi^0$. The
width was predicted in Ref.~\cite{Guo:2010ak} as $(0.29\pm0.10) (g_1 g_1'~{\rm
GeV})^2~{\rm keV}$, see footnote~\ref{foot1}. The right panel of
Fig.~\ref{fig:hcM1} shows the parameter-free predictions of the ratios
$\Gamma(h_c'\to\gamma\chi_{cJ})/\Gamma(\chi_{c2}'\to\pi^0\chi_{c1})$ as a function
of the mass of the $h_c'$. Such predictions can only be made when both processes
are loop-dominated because only in this case the decay amplitudes are proportional
to the same product of coupling constants $g_1 g_1'$. Were they
multipole-dominated, an unknown matrix element of gluon operators would be involved
in the transition $\chi_{c2}'\to\pi^0\chi_{c1}$ so that the process cannot be
directly related to the hindered M1 transitions. Even though the uncertainty of
these predictions is sizeable, they are markedly different from potential model
calculations.

In this Letter, we argue that the hindered M1 transitions of $P$-wave charmonium
are dominated by the coupled-channel effects. The conclusion is supported by
numerical calculations. With a reasonable estimate of the unknown coupling
constants, the results turn out to be much larger than those obtained in the quark
model. Parameter-free predictions are made for ratios of partial widths of two
completely different types of charmonium transitions: the hindered M1 transitions
and single-pion transitions of the $P$-wave charmonia. The $P$-wave charmonia
considered here are assumed to be $c\bar c$ states so that they are organized as in
Eq.~(\ref{eq:chicJ}). If their coupling to the charmed mesons becomes resonant,
renormalization is necessary (see, e.g.
Refs.~\cite{Braaten:2007nq,Artoisenet:2010va}). In that case, the resulting widths
for the hindered M1 transitions should still be much larger than the results were
the transitions not dominated by the coupled-channel effects, since a resonant
coupling tends to enhance the widths further. Experimental efforts on measuring the
transitions suggested here are needed towards understanding the coupled-channel
effects in the charmonium transitions. In fact, the existing technologies of
lattice calculations of the heavy quarkonia radiative
transitions~\cite{Dudek:2009kk,Chen:2011kp,Lewis:2011ti} are well ready to test the
picture presented in this Letter: if the hindered M1 transitions of the $P$-wave
charmonia are dominated by the coupled-channel effects, the results of simulations
with dynamical light quarks should be significantly larger than those in the
quenched approximation.

We further notice that the hindered M1 transitions of the $S$-wave charmonia are
not well suited for studying the coupled-channel effects. This is because for these
transitions the charmed meson loops are divergent. While the divergence of the
triangle diagrams scales as $\order[v]{0}$ in the velocity counting, the finite
part scales as $\order[v]{}$. The divergence must be absorbed by a counterterm.
However, the counterterm cannot be determined from elsewhere. Hence, parameter-free
predictions analogous to those made in this Letter is not possible.

\begin{acknowledgments}
We thank Christoph Hanhart for useful discussions. This work is partially supported
by DFG through funds provided to the SFB/TR 16 and the EU I3HP ``Study of Strongly
Interacting Matter'' under the Seventh Framework Program of the EU. U.-G. M. also
thanks the BMBF for support (Grant No. 06BN9006). F.-K. G. also thanks partial
support from the NSFC (Grant No. 11165005).
\end{acknowledgments}

\end{document}